# Finite-time Stability Analysis for Random Nonlinear Systems

Sina Sanjari, Mahdieh Tahmasebi*

*Abstract*— This paper presents an analysis approach to finite-time attraction in probability concerns with nonlinear systems described by nonlinear random differential equations (RDE). RDE provide meticulous physical interpreted models for some applications contain stochastic disturbance. The existence and the path-wise uniqueness of the finite-time solution are investigated through nonrestrictive assumptions. Then a finite-time attraction analysis is considered through the definition of the stochastic settling time function and a Lyapunov based approach. A Lyapunov theorem provides sufficient conditions to guarantee finite-time attraction in probability of random nonlinear systems. A Lyapunov function ensures stability in probability and a finiteness of the expectation of the stochastic settling time function. Results are demonstrated employing the method for two examples to show potential of the proposed technique.

## I. INTRODUCTION

Stability has been a subject of matter of the numbers of articles in the control theory and engineering (see for instance [1, 2] and their references). Several articles have developed different methods to address the stability criteria for deterministic and stochastic systems. Stability of stochastic differential equations (SDE) was investigated by literatures involves stochastic stability in probability, almost sure stability, and etc. (see for example [3]). For several practical applications a particular property of stability is required called "finite-time attraction" to ensure finite-time attractiveness of the equilibrium [4, 5]. Several applications such as a secure communication [6], a robot manipulator [4], a sliding mode controller, and a sliding mode observer require a finite-time property to verify the results deal with either stochastic or deterministic systems [7-9]. The concept of finite-time stability was investigated in associated with a Lyapunov technique in [10]. A finite-time stability analysis was extended to stochastic systems described by stochastic differential equations in [11] by defining the concept of a stochastic settling time function. Moreover, finite-time stability in probability was characterized by Lyapunov constraints to ensure the existence of the finite stochastic settling time function [12].

Stochastic differential equations have been proven to be an appropriate model to fit the data in many applications, but in some situations they provide inappropriate model to describe systems contain stochastic disturbance and therefore suffer from some disadvantages [13]. For example, white noise is driven from the wiener process whose does not have derivative everywhere, so it is unsuitable to model fluctuation in practical applications. Furthermore, The Hessian term that exists in the Itô formula is difficult to interpret physically, and is hard to handle for stability analysis. There is also a difficulty in the selection between two well-known descriptions of SDE, an Itô integral equation and a Stratonovich integral equation, for a specific application. Moreover, since a white noise is unbounded, it fails to describe the model of some applications; therefore, other stochastic processes such as a stationary process are required; For instance, modeling road irregularity impact on an operation system of cars, and modeling circuit systems contain a noise filter, and several applications in an automatic control, an information theory, and a wireless technology [14, 15]. Because of these limitations and difficulties, SDE model is not accurate enough to model all the applications contain a stochastic disturbance. To address this problem nonlinear random models have been acquired to alleviate the aforementioned limitations [13]. Furthermore, it was shown that in dealing with random nonlinear models we can propose more practical solution for a regulation problem that is independent from the reference signals [13]. Moreover, nonlinear random model enables some deterministic analysis tools to be applied [14]; while model described by stochastic differential equations does not have this potential, because of the existence of the second order Euler approximation term [2]. Stability results of the nonlinear RDE have been presented in [16], but some restrictive assumptions were acquired to conclude the existence and uniqueness of a stable solution [17]. These restrictive assumptions and constraints confine the extension of RDE to the wide range of applications dealing with a stochastic disturbance. However, recently, [13] construct a general framework to address the stability criteria of nonlinear RDE, and presents theorems employing mild assumptions that conclude the stability of RDE based on a Lyapunov approach, which renders extending the applications of nonlinear RDE especially in the control theory [15, 18]. Noise to state stability, asymptotic gain properties, global

Sina Sanjari is with Tarbiat Modares University, Tehran, IRAN (e-mail: s.sanjari@modares.ac.ir).
Mahdieh Tahmasebi is with Tarbiat Modares University, Tehran, IRAN. (e-mail: tahmasebi@modares.ac.ir).

asymptotic stability were all investigated for RDE in [13, 15]. However, to the best of authors' knowledge no work on finite-time stability of nonlinear RDE has been developed until now.

In this paper, we mainly deal with finite-time attraction of random nonlinear systems. We present sufficient conditions based on a Lyapunov approach to ensure finite-time attraction of the equilibrium. First, we introduce unrestrictive assumptions to guarantee the path-wise uniqueness and the existence of the finite-time solution. Then along with the satisfaction of the weak law of large numbers assumption on random process, we proposed finite-time stability satisfying inequality constraints on the Lyapunov function. These constraints provide a general framework to analysis finite-time attraction in probability of the system trajectories that ensure the existence and the finiteness of the stochastic settling function, which is defined as an expectation of the settling time function.

The rest of the paper is organized as follows: Section II presents the path-wise uniqueness assumption and the required conditions for the existence of the finite-time solution. Then, in section III, based on the assumptions in section II, we present a theorem and corollary to establish finite-time stability concerns with random nonlinear systems using the Lyapunov approach. In section IV, numerical examples are given to demonstrate the effectiveness of the presented analysis. Finally, section V concludes the paper.

**Notation**: Real $n$-dimensional space denotes by $\mathbb{R}^n$, and set of all nonnegative real numbers $\mathbb{R}_+$; $U_R$ stands for the ball $|x| < R$. For a vector $x$, $|x|$ stands for Euclidean norm, and real $n$-dimensional space denotes by $\mathbb{R}^n$, $\|Q\|$ is the 2-norm of a matrix $Q$. $C^i$ stands for the set of all functions with continuous $i$-th partial derivative; set of all continuous strictly increasing and vanish at zero functions $\mathbb{R}_+ \to \mathbb{R}_+$ denotes by $\mathcal{K}$, and a set of all unbounded functions in class-$\mathcal{K}$ denotes by $\mathcal{K}_\infty$. $\mathcal{KL}$ stands for the set of all functions $\beta(s,t): \mathbb{R}_+ \times \mathbb{R}_+ \to \mathbb{R}_+$ which is of class-$\mathcal{K}$ for each fixed $t$, and decreases to zero as $t \to \infty$ for each fixed $s$.

## II. SYSTEM DESCRIPTION AND PROBLEM STATEMENT

Consider the following nonlinear random system

$$\dot{x} = f(x(t), t) + g(x(t), t)\xi(t) \qquad (1)$$

Where $x \in \mathbb{R}^n$ is the state vector, $\xi \in \mathbb{R}^l$ is $\mathcal{F}_t$-adapted and piecewise continuous stochastic process, $f(x,t): \mathbb{R}^n \times \mathbb{R}_+ \to \mathbb{R}^n$ is a known nonlinear function, and $g(x,t): \mathbb{R}^n \times \mathbb{R}_+ \to \mathbb{R}^{n \times l}$ is a known full rank state-dependent matrix.

**Assumption 1**: The stochastic process $\xi(t)$ is $\mathcal{F}_t$-adapted, piecewise continuous such that there exists a positive constant $K$ satisfies

$$\sup_{t \geq t_0} E\{|\xi(t)|^2\} < K \qquad (2)$$

It means that the mean-square of the stochastic process $\xi(t)$ is bounded by a constant.

**Assumption 2**: The solution $x(t)$ of (1) is continuous, $\mathcal{F}_t$-adapted, and satisfies $\forall t \in [t_0, T]$

$$x(t) = x(t_0) + \int_{t_0}^T f(x,s)ds + \int_{t_0}^T g(x,s)\xi(s)ds$$

**Assumption 3**: Nonlinear functions $f(\cdot,\cdot), g(\cdot,\cdot)$ vanish at the origin, i.e., $f(0,t) = g(0,t) = 0 \ \forall t \in [t_0, \infty)$.

**Lemma 1**: Under the above assumptions, if the following two conditions hold for continuous functions $f(x,t), g(x,t)$ for each $t \in [t_0, T]$

$$|f(x_1(t), t) - f(x_2(t), t)| \leq c_1(t)\kappa(|x_1(t) - x_2(t)|) \qquad (3)$$

Where $c_1(t)$ is nonnegative function such that $\int_0^T c(t)dt < \infty$ and $\kappa(u) \geq 0$ for $u \geq 0$ is a deterministic, strictly increasing, continuous concave function satisfies:

$$\|g(x_1, t) - g(x_2, t)\|^2 \leq c_2(t)\rho(|x_1 - x_2|) \qquad (4)$$

Where $c_2(t)$ is nonnegative function such that $\int_0^T c_2(t)dt < \infty$ and $\rho(u) \geq 0$ for $u \geq 0$ is a deterministic, strictly increasing, continuous concave function satisfies $\int_{0^+}^\gamma du/\rho(u) = +\infty$ and $\int_{0^+}^\gamma du/(\sqrt{\rho(u)} + \kappa(u)) = +\infty$ for each $\gamma > 0$. Then, for any given $x(t_0) \in \mathbb{R}^n$, (1) has a path-wise strong unique solution.

**Proof:** Let $x_1(t)$ and $x_2(t)$ be two solutions of (1) where $x_1(t_0) = x_2(t_0)$. According to the assumption 2, we can write

$$x_1(t) - x_2(t) = \int_{t_0}^t (f(x_1(s), s) - f(x_2(s), s) + (g(x_1(t), s) - g(x_2(s), s))\xi(s))ds$$

According to the Euler formula,

$$|x_1(t) - x_2(t)| \leq \int_{t_0}^t (|f(x_1(s), s) - f(x_2(s), s)| + (\|g(x_1(s), s) - g(x_2(s), s)\|)|\xi(s)|)ds$$

Taking expectation from both sides implies that

$$E|x_1(t) - x_2(t)| \leq E\int_{t_0}^t c_1(s)\kappa(|x_1(s) - x_2(s)|)ds + E\int_{t_0}^t c_2(s)\|g(x_1(s), s) - g(x_2(s), s)\| |\xi(s)|ds := E\{I_1\} + E\{I_2\}$$

From Fubini's theorem and since $c_1(s)$ is deterministic, we deduce that

$$E\{I_1\} \leq \int_{t_0}^t c_1(s)E\kappa(|x_1(s) - x_2(s)|)ds$$

Since $\kappa(\cdot)$ is concave and positive, Jensen's inequality implies that

$$E\{I_1\} \leq \int_{t_0}^t c_1(s)\kappa(E|x_1(s) - x_2(s)|)ds$$

Utilizing Cauchy-Schwarz inequality for $I_2$ implies that

$$E\{I_2\} \leq \int_{t_0}^{t} c_2(s)(E\|g(x_1(s),s) - g(x_2(s),s)\|^2)^{\frac{1}{2}}(E|\xi(s)|^2)^{\frac{1}{2}} ds$$

(2) yields that

$$E\{I_2\} \leq \sqrt{K} \int_{t_0}^{t} c_2(s)(E\rho(|x_1(s) - x_2(s)|))^{\frac{1}{2}} ds$$

Following the concavity of $\rho(\cdot)$, we have

$$E\{I_2\} \leq \int_{t_0}^{t} c_2(s)(\rho(E|x_1(s) - x_2(s)|))^{\frac{1}{2}} ds$$

Then

$$E|x_1(t) - x_2(t)| \leq \int_{t_0}^{t} c(s)\rho'(E|x_1(s) - x_2(s)|) ds$$

Where $\rho'(\cdot) = \sqrt{K\rho(\cdot)} + \kappa(\cdot)$ is a concave function since $\kappa$ and $\rho$ are strictly increasing and concave functions. As it is well-known [19, 20] this equals to $E|x_1(t) - x_2(t)| = 0$, so $x_1(t) = x_2(t)$. The proof is completed.
∎

**Assumption 4**: $|\xi(t)|^2$ satisfies a weak law of large numbers i.e., For any sufficiently small positive constants $\varepsilon$ and $\delta$, there exists a positive constant $T_1 > t_0$ such that for all $t \geq T_1$

$$P\{\left|\frac{1}{t-t_0}\int_{t_0}^{t}|\xi(s)|^2 ds - E\{|\xi(t)|^2\}\right| \geq \delta\} \leq \varepsilon \quad (5)$$

It is easily proven using Chebyshev's inequality. ∎

**Remark 1**: The above assumptions on stochastic processes $\xi(t)$ are reasonable. Several stochastic processes were introduced in [13] verified in these assumptions; for instance, a second-order process, a widely periodic process, a widely stationary process, a strictly stationary process, and a stationary Gaussian process. ∎

**Definition 1**: The equilibrium $x(t) \equiv 0$ of the system (1) is
1) Globally stable in probability if for every positive constant $\varepsilon$, there exists a class-$\mathcal{K}$ function $\gamma(\cdot)$ such that $\forall t \geq t_0, x(t_0) \in \mathbb{R}^n \setminus \{0\}$

$$P\{|x(t,x(t_0))| \leq \gamma(|x(t_0)|)\} \geq 1 - \varepsilon \quad (6)$$

2) Globally asymptotically stable in probability if $\forall \varepsilon > 0$, there exists class-$\mathcal{KL}$ function $\beta(\cdot,\cdot)$ such that

$$P\{|x(t,x(t_0))| \leq \beta(|x(t_0)|, t - t_0)\} \geq 1 - \varepsilon \quad (7)$$
∎

Now we define finite-time stability in probability for the systems described by random nonlinear differential equations [13], according to the definition of this kind of stability for the systems described by stochastic differential equations [11]. First, we define the stochastic settling time function, and then we present conditions guarantee finite-time stability.

**Definition 2**: The function $T_0(x(t_0), w) = \inf\{T \geq 0; x(t,x(t_0)) = 0, \forall t \geq T\}$ is called stochastic settling function for (1). The stochastic settling time function $T_0(x(t_0), w)$ emphasizes that the random nonlinear system will reach to the origin in a finite time.

**Definition 3:** The equilibrium $x(t) \equiv 0$ of (1) is globally finite-time stable in probability, if for every $x(t_0) \in \mathbb{R}^n$ two following conditions hold.
(I) Stochastic settling function $T_0(x(t_0), w)$ exists with probability one.

(II) $P\{T_0 < \infty\} = 1$.

The main goal of this technical note is to provide conditions based on Lyapunov function to ensure finite-time attraction in probability for the random nonlinear systems.

### III. MAIN RESULTS

This section presents a theorem to establish a Lyapunov function that ensures finite-time stability for the system introduced in section II along with the assumptions required in this section.

**Theorem 1:** Under Assumptions 1-4, there exists a unique solution of (1) on $[t_0, \infty)$ in which its equilibrium $x(t) \equiv 0$ is globally finite-time stable in probability if there exist a $C^1$ function $V: \mathbb{R}^n \to \mathbb{R}_+$ satisfying $\lim_{k \to \infty} \inf_{|x|>k} V(x(t)) = \infty$ and a function $r: \mathbb{R}_+ \to \mathbb{R}_+$, and class-$\mathcal{K}_\infty$ functions $\alpha_1, \alpha_2$ and constants $c_1, c_2 > 0$, $c_1 > 2c_2\sqrt{K}$ such that

$$\alpha_1(|x(t)|) \leq V(x(t)) \leq \alpha_2(|x(t)|) \quad (8)$$

$$\frac{\partial V(x)}{\partial x} f(x,t) \leq -c_1 r(V(x)), \left|\frac{\partial V(x)}{\partial x} g(x,t)\right| \leq c_2 r(V(x)) \quad (9)$$

$$\text{for every } 0 \leq \varepsilon < +\infty, \int_0^\varepsilon \frac{1}{r(v)} dv < +\infty \quad (10)$$

Moreover, the settling time function $T_0(x(t_0), w)$ satisfies

$$E\{T_0(x(t_0), w)\} \leq \frac{1}{(c_1 - 2c_2\sqrt{K})} \int_0^{V(x(t_0))} \frac{1}{r(v)} dv$$

**Proof:** To prove the theorem we first show that the origin of the system is globally asymptotically stable in probability, that concludes that for every $x(t_0) \in \mathbb{R}^n$, the settling time function $T_0(x(t_0), w)$ exists with probability one. Next, we demonstrate that $E\{T_0(x(t_0), w)\} < \infty$ according to the conditions in Theorem 1. Finally, we complete the proof using Definition 2.

*Step 1*. Define set $U_k = \{x; |x| < k\}$ and assume that there exists escaping time $\sigma_k = \inf\{t \geq t_0: |x(t)| \geq k\}$. Define the limit of that as $\sigma_\infty = \lim_{k \to \infty} \sigma_k$. According to Lemma 1, it can be easily verified that there exist a path-wise unique maximal solution $\{x(t): t_0 \leq t < \sigma_\infty\}$ of (1) for $[t_0, \infty)$ and $k > 0$ almost sure such that

$$x(t \wedge \sigma_k) = x(t_0) + \int_{t_0}^{t \wedge \sigma_k} f(x(s), s) ds + \int_{t_0}^{t \wedge \sigma_k} g(x(s), s)\xi(s) ds$$

For the existence of the global unique maximal solution we require to prove that $\sigma_\infty = \infty$ almost surely.

Consider $V(x(t))$ as a Lyapunov candidate function. $V(x(t))$ is positive definite from (7), and the derivative of $V(x(t))$ along with (1) and (9) implies that

$$\dot{V}(x(t \wedge \sigma_k)) = \frac{\partial V(x(t \wedge \sigma_k))}{\partial x} f(x(t \wedge \sigma_k), t)$$
$$+ \frac{\partial V(x(t \wedge \sigma_k))}{\partial x} g(x(t \wedge \sigma_k), t)\xi(t)$$
$$\leq (-c_1 + c_2|\xi(t)|) r(V(x(t \wedge \sigma_k)))$$

So,

$$\frac{\partial V(x(t \wedge \sigma_k))}{r(V(x(t \wedge \sigma_k)))} \leq (-c_1 + c_2|\xi(t)|) dt \quad (11)$$

(10) implies that $\theta(V(x(t \wedge \sigma_k))) := \int_0^{V(x(t \wedge \sigma_k))} \frac{1}{r(v)} dv$ exists, so

$$\int_{t_0}^t \frac{\partial V(x(s \wedge \sigma_k))}{r(V(x(s \wedge \sigma_k)))} \leq \int_{t_0}^t (-c_1 + c_2|\xi(s)|) ds$$
$$\int_{t_0}^t \frac{\partial \theta(V(x(t \wedge \sigma_k)))}{\partial V} \partial(V(x)) = \int_{t_0}^t \partial \theta(V(x(t)))$$
$$\leq \int_{t_0}^t (-c_1 + c_2|\xi(s)|) ds$$

$$\theta(V(x(t \wedge \sigma_k))) \leq \theta(V(x(t_0)) + \int_{t_0}^t (-c_1 + c_2|\xi(s)|) ds \quad (12)$$

Taking expectation implies that

$$E\theta(V(x(t \wedge \sigma_k))) \leq \theta(V(x(t_0)) + E \int_{t_0}^t (-c_1 + c_2|\xi(s)|) ds$$

Let us to define the set $A$ as the following form for each sufficiently small $\varepsilon > 0$ and $\delta \in (0, 3K)$

$$A = \left\{ \left| \frac{1}{t-t_0} \int_{t_0}^t |\xi(s)|^2 ds - E\{|\xi(t)|^2\} \right| \leq \delta \right\}$$

From Assumption 4, there exists $T_1 > 0$ such that for all $t \geq T_1$, (5) holds. Based on assumption 1 and 4, for all $w \in A$

$$\int_{t_0}^t |\xi(s)|^2 ds \leq (t - t_0)(E\{|\xi(t)|^2\} + \delta) \leq 4K(t - t_0),$$
$$\forall k > 0$$

and then

$$\int_{t_0}^t |\xi(t)| dt \leq \sqrt{(t-t_0)} \left( \int_{t_0}^t |\xi(t)|^2 dt \right)^{\frac{1}{2}} \leq 2\sqrt{K}(t - t_0) \quad (13)$$

Therefore we can derive that

$$E\theta(V(x(t \wedge \sigma_k))) \leq \theta(V(x(t_0))e^{ct}, \forall k > 0 \quad (14)$$

where $c$ is a positive constant.
Now, suppose $\sigma_\infty < \infty$ a.s., so

$$P\{\sigma_\infty \leq T\} > 2\varepsilon$$

By definition of $\sigma_\infty$, there exists integer $k \geq k_0$ such that

$$P\{\sigma_k \leq T\} > \varepsilon \quad (15)$$

For any $t_0 \leq t \leq T$, from (14) we derive that

$$E\{I_{\sigma_k \leq T} \theta(V(x(\sigma_k))) \leq de^{ct}, \forall k > 0 \quad (16)$$

Where $d$ is a positive constant. On the other hand, we define

$$h_k = \inf\{V(x(t)): |x| \geq k, t \in [t_0, T]\}$$

According to the definition of $\theta(V)$ and assumptions on $V$, we conclude that $h_k \to \infty$ as $k \to \infty$.
On the other hand (15) and (16) implies that

$$de^{ct} \geq h_k P\{\sigma_k \leq T\} > \varepsilon h_k$$

Letting $k \to \infty$ leads to contradiction, so we can deduce that the explosion time $\sigma_\infty = \infty, a.s.$. Therefore, there exists a global unique maximal solution $\{x(t): t_0 \leq t < \infty\}$ of (1). Thus,

$$\theta(V(x(t))) \leq \theta(V(x(t_0)) + \int_{t_0}^t (-c_1 + c_2|\xi(s)|) ds \quad (17)$$

Substituting (13) into (17), along with (10), we derive

$$V(x(t)) \leq \theta^{-1}\{\theta(V(x(t_0))) - (c_1 - 2c_2\sqrt{K})(t - t_0)\}$$

Regarding to (8) and (5), we can concludes that

$$P\{|x(t)| \leq \alpha_1^{-1}(\theta^{-1}\{\theta(\alpha_2(|x(t_0)|)) - (c_1 - 2c_2\sqrt{K})(t - t_0)\}\} \geq 1 - \varepsilon, \quad \forall t \geq T_1 \quad (18)$$

It is straightforward from (18) and (8) that we can prove that there exists a class-$\mathcal{KL}$ function $\beta_1(\cdot, \cdot)$ such that (7) holds. On the other hand for $t \leq T_1$, from Markov inequality, we have

$$P\{|\xi(t)| > \delta_0\} \leq E|\xi(t)|^2/\delta_0^2 \leq K/\delta_0^2 = \varepsilon, \forall t \leq T_1$$

So,

$$V(x(t)) \leq \theta^{-1}\{\theta(V(x(t_0))) - (c_1 - c_2\delta_0)(T_1 - t_0)\}$$

Therefore,

$$P\{|x(t)| \leq \beta_2(|x(t_0)|)\} \geq 1 - \varepsilon, \quad \forall t \leq T_1$$

where $\beta_2(\cdot, \cdot)$ is a class-$\mathcal{KL}$ function. Thus, from the definition of asymptotic stability we can conclude that the origin of (1) is globally asymptotic stable in probability; Therefore, the settling time function $T_0(x_0, w)$ exists with probability one.
*Step 2.* Now, we prove that $E\{T_0(x_0, w)\} < \infty$, and exists.
According to (11) we have

$$\int_0^{T_0} \frac{\partial V}{r(V(x))} \leq \int_0^{T_0} (-c_1 + c_2|\xi(t)|) dt$$

And then

$$\int_0^{T_0} \frac{\partial \theta(V)}{\partial V} \partial(V(x)) \leq \int_0^{T_0}(-c_1 + c_2|\xi(t)|)dt$$

Taking expectation from both sides, and using (13), we derive

$$E\left\{\int_0^{T_0} \partial\theta(V)\right\} \leq -c_1 E\{T_0(x(t_0), w)\} + c_2 E\left\{\int_0^{T_0}|\xi(t)|dt\right\}$$

$$E\{\theta(V)|_0^{T_0}\} \leq -c_1 E\{T_0(x(t_0), w)\} + c_2 E\{\int_0^{T_0}|\xi(t)|dt\} \quad (19)$$

Since $\theta(V(x(T_0))) = 0$, then

$$-E\{\int_0^{V(x(t_0))} \frac{1}{r(v)} dv\} \leq -(c_1 - 2c_2\sqrt{K})E\{T_0\}$$

Therefore,

$$E\{T_0\} \leq \frac{1}{(c_1 - 2c_2\sqrt{K})} E\{\int_0^{V(x(t_0))} \frac{1}{r(v)} dv\}$$
$$= \frac{1}{(c_1 - 2c_2\sqrt{K})} \int_0^{V(x_0)} \frac{1}{r(v)} dv < +\infty$$

It means that $E\{T_0\}$ exists and is finite. From Markov inequality for all $\varepsilon_1 > 0$ we can deduce that

$$P\{T_0 \geq \varepsilon_1\} \leq E\{T_0\}/\varepsilon_1$$

Therefore,

$$P\{T_0 < \varepsilon_1\} = 1 - P\{T_0 \geq \varepsilon_1\} \geq 1 - E\{T_0\}/\varepsilon_1$$

Since $E\{T_0\}$ is finite, for $\varepsilon_1 \to \infty$ we can deduce that

$$P\{T_0 < \infty\} = 1$$

Which completes the proof. ∎

**Remark 2:** Lemma 1 gives unrestrictive condition to guarantee the existence and uniqueness of the solution of (1). This unrestrictive condition is required for discussing the finite time stability since conditions of Theorem 1 and results of theorems in [21] conclude that there exists coefficient of (1) that fails to satisfy local Lipchitz condition, if the solution of (1) be finite time stable in probability. ∎

**Corollary 1:** Under Assumptions 1-4, there exists a unique solution of system (1) on $[t_0, \infty)$, and its equilibrium $x(t) \equiv 0$ is globally finite-time stable in probability if there exist a $C^1$ function $V: \mathbb{R}^n \to \mathbb{R}_+$, satisfying $\lim_{k\to\infty} \inf_{|x|>k} V(x(t)) = \infty$, and class-$\mathcal{K}_\infty$ functions $\alpha_1, \alpha_2$, and constants $c_1, c_2 > 0$, $c_1 > 2c_2\sqrt{K}$ and $0 \leq \gamma < 1$ such that

$$\alpha_1(|x(t)|) \leq V(x(t)) \leq \alpha_2(|x(t)|) \quad (20)$$
$$\frac{\partial V}{\partial x} f(x,t) \leq -c_1 V(x)^\gamma, \left|\frac{\partial V}{\partial x} g(x,t)\right| \leq c_2 V(x)^\gamma \quad (21)$$

Then the stochastic settling time function $T_0(x(t_0), w)$ satisfies $E\{T_0(x(t_0), w)\} \leq \frac{(V(x_0))^{1-\gamma}}{(1-\gamma)(c_1-2c_2\sqrt{K})}$

**Proof:** Take $r(V) = V(x)^\gamma$, then (21) implies that (9) and (10)

hold. The rest of the proof is similar to the proof of Theorem 1. ∎

## IV. ILLUSTRATIVE EXAMPLES

In this section, some examples are provided to show the effectiveness of the analysis developed in this paper.

**Example 1:** Consider the following nonlinear random system with stochastic disturbance.

$$\begin{cases} \dot{x}_1 = -x_1^{\frac{1}{3}} - \frac{1}{2}x_1 + \frac{2}{3}x_2 + x_1^{\frac{1}{3}}\xi_1(t) \\ \dot{x}_2 = -x_2^{\frac{1}{3}} - x_2 + \frac{1}{3}x_1 + x_2^{\frac{1}{3}}\xi_2(t) \end{cases} \quad (22)$$

Where $\xi_1(t), \xi_2(t)$ are stochastic processes satisfy Assumptions 1-4. It is straightforward to verify that (22) admits the existence and path-wise uniqueness of the solution. In the simulation, we consider $\xi_1(t), \xi_2(t) \in \mathbb{R}$ for $t \in T$ as a stationary process. By considering a Lyapunov function in Corollary 1, $V(x) = \frac{1}{2}(x_1^2 + x_2^2)$, so

$$\frac{\partial V}{\partial x} f(x,t) = -x_1^{\frac{4}{3}} - x_2^{\frac{4}{3}} - \frac{1}{2}x_1^2 - x_2^2 + x_1 x_2$$

applying Young's inequality and using an elementary inequality in [22] gives that

$$\frac{\partial V}{\partial x} f(x,t) \leq -x_1^{\frac{4}{3}} - x_2^{\frac{4}{3}} \leq -V(x)^\gamma, \quad 0 < \gamma < 1$$

It is easy to show that $\left|\frac{\partial V}{\partial x} g(x,t)\right| \leq c_2 V(x)^\gamma$.

Fig.1 shows the state response of (22), which demonstrates finite-time attraction of the solution.

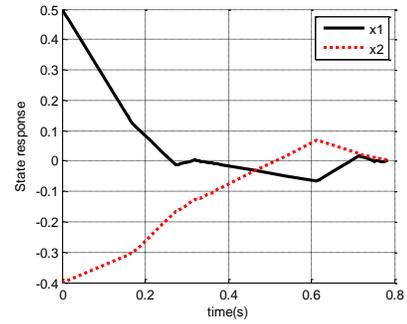

**Fig 1.** Response random system of example 1

In the simulation, we consider

$$\xi(t) = -K\cos(wt + S)$$

Where $K$ and $w$ are real positive constants, and $S$ is a random variable with the uniform distribution on interval $[0, 2\pi]$. Employed random process holds assumptions in section II.

**Example 2:** Consider a scalar system with stochastic disturbance as follows:

$$\dot{x} = 3x\frac{(arctanx)^2}{(1+x^2)} + u - \frac{x}{(1+x^2)} + \frac{1}{2}(1+x^2)(arctanx)^{\frac{1}{3}}\xi(t)$$

Where $x \in \mathbb{R}$ is state, and $u \in \mathbb{R}$ is control input, and $\xi(t)$ is piecewise continuous adapted stochastic process that satisfies required assumptions in section II.

The control objective is to design control signal $u$ such that the closed-loop dynamics became finite-time stable in probability. To this end, we consider the following Lyapunov candidate function

$$V(x) = \frac{1}{2}(arctanx)^2$$

So,

$$\frac{\partial V}{\partial x}f(x) = \frac{arctanx}{1+x^2}(3x\frac{(arctanx)^2}{(1+x^2)} + u - \frac{1}{(1+x^2)})$$

Choose

$$u(x) = -(1+x^2)(arctanx)^{\frac{1}{3}} - 3x\frac{(arctanx)^{2-\frac{1}{3}}}{(1+x^2)}$$

Then,

$$\frac{\partial V}{\partial x}f(x) = -(arctanx)^{\frac{4}{3}} = -V(x)^{\frac{1}{3}}$$

It is not difficult to show that $\left|\frac{\partial V}{\partial x}g(x)\right| \leq \frac{1}{2}V(x)^{\frac{1}{3}}$. So regarding to Corollary 1 the closed-loop dynamics is finite-time stable. Fig.2 shows the response of the closed-loop dynamics.

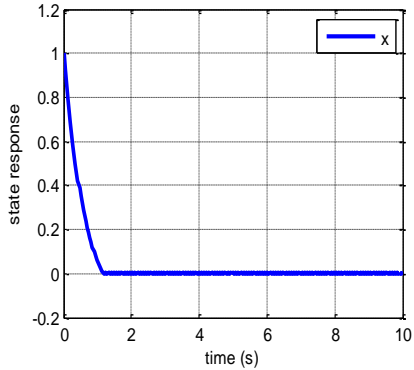

**Fig 2**. Response random system of example 2

For simulation purpose, $\xi(t)$ is considered as a zero-mean widely stationary process with mean square value $A/2r$:

$$r\dot{\xi}(t) = -\xi(t) + w(t)$$

Where $\xi(0) = 0$, $r > 0$, and $w(t)$ is a zero-mean white noise, The power spectral of $\xi(t)$ is

$$F_\xi(\lambda j) = \begin{cases} \frac{A}{1+r^2\lambda^2} & |\lambda| \leq \lambda_c \\ 0 & otherwise \end{cases}$$

Fig.3 demonstrates the control input signal, and the stochastic disturbance.

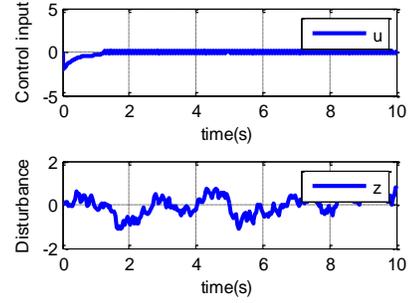

**Fig 3**. Control input, and the stochastic disturbance of example 2.

## V. CONCLUSION

In this technical note, we introduce a general framework to analysis finite-time attraction in probability for random nonlinear systems. A Lyapunov theorem was presented to conclude finite-time stability for a particular type of stochastic systems described by random nonlinear differential equations. We presented new unrestrictive and reasonable assumptions to ensure the existence and uniqueness of the finite-time solution. Several examples were presented to verify an applicability of the proposed method. One outstanding Benefits of this analysis is presenting a general theoretical framework for finite-time attraction in probability for nonlinear random systems that enables future analysis in designing an optimal control and the sliding mode control. In addition, the results can be extended for a constructive design method of a controller ensures finite-time attraction in probability of random nonlinear systems. Further analysis will be required to extend the results to switching systems described by random nonlinear systems.